\documentclass[aps,prc,twocolumn,showpacs,superscriptaddress,footinbib]{revtex4-1}

\usepackage{graphicx}
\usepackage{dcolumn}
\usepackage{bm}
\usepackage{enumerate}
\usepackage{amsmath}

\usepackage[normalem]{ulem}  
\usepackage[colorlinks]{hyperref}

\begin{document}


\title{Post breakup dynamics of fragments produced in nuclear multifragmentation}

\author{S.R. Souza}
\affiliation{Instituto de F\'\i sica, Universidade Federal do Rio de Janeiro Cidade Universit\'aria, 
\\Caixa Postal 68528, 21941-972 Rio de Janeiro-RJ, Brazil}
\affiliation{Departamento de F\'\i sica, ICEx, Universidade Federal de Minas Gerais,
\\Av.\ Ant\^onio Carlos, 6627, 31270-901 Belo Horizonte-MG, Brazil}
\author{B.V. Carlson}
\affiliation{Departamento de F\'\i sica, Instituto Tecnol\'ogico de Aeron\'autica-CTA, 12228-900 S\~ao Jos\'e dos Campos-SP, Brazil}
\author{R. Donangelo}
\affiliation{Instituto de F\'\i sica, Universidade Federal do Rio de Janeiro Cidade Universit\'aria, 
\\Caixa Postal 68528, 21941-972 Rio de Janeiro-RJ, Brazil}
\affiliation{Instituto de F\'\i sica, Facultad de Ingenier\'\i a, Universidad de la Rep\'ublica, 
Julio Herrera y Reissig 565, 11.300 Montevideo, Uruguay}

\date{\today}

\begin{abstract}
The deexcitation of the primary hot fragments, produced in the breakup of an excited nuclear source, during their propagation under the influence of their mutual Coulomb repulsion is studied in the framework of a recently developed hybrid model.
The latter is based on the Statistical Multifragmentation Model (SMM), describing the prompt breakup of the source, whereas the particle emission from the hot fragments, that decay while traveling away from each other, is treated by the Weisskopf-Ewing evaporation model.
Since this treatment provides an event by event description of the process, in which the classical trajectories of the fragments are followed using molecular dynamics techniques, it allows one to study observables such as two-particle correlations and infer the extent to which the corresponding observables may provide information on the multifragment production mechanisms.
Our results suggest that the framework on which these treatments are based may be considerably constrained by such analyses.
Furthermore, they imply that information obtained from these model calculations may provide feedback to the theory of nuclear interferometry.
We also found that neutron deficient fragments should hold information more closely related to the breakup region than neutron rich ones, as they are produced in much earlier stages of the post breakup dynamics than the latter.

\end{abstract}

\maketitle

\begin{section}{Introduction}
\label{sect:introduction}
The understanding of the underlying physics governing the copious production of fragments, as a result of not-too-peripheral collisions between heavy ions at energies above a few tens of MeV per nucleon, is a challenging task both from the theoretical and the experimental point of view \cite{BorderiePhaseTransition2008,BettyPhysRep2005,reviewSubal2001}.

Theoretical treatments should, ideally, consider quantum many-body effects in a strongly interacting fermionic system that reaches stages far from equilibrium.
At this point, the nuclear matter is already hot and compressed  and pre-equilibrium emission starts, prior to a fast expansion that may drive the system to much lower densities \cite{timeScaleExpansionEmission1995,exoticDens,FlowSchussler,AMDReview2004,timeScaleDynStatistical2006,BonaseraBUU}.
Since such a complete a treatment is beyond the present computational capabilities, several theoretical models, based on simplifying assumptions, have been proposed, focusing on different aspects of the collision process.
Dynamical models \cite{AMDReview2004,timeScaleDynStatistical2006,FMD2000,BonaseraBUU,AichelinPhysRep,BurgioSpinodalInstabilities1994,constrainedMolecularDynamics} are the most ambitious ones as they generally aim to describe the collision from the very early phase when the nuclei are far away from each other to the late stages after the hot fragments have been created.
Assuming that thermal equilibrium is attained after the most violent stages of the reaction, statistical models take the configuration of the decaying source for granted and, from it, predict the properties of the hot primary fragment distribution \cite{ReviewIgor2006,BettyPhysRep2005,GrossPhysRep,Bondorf1995}.
This hypothesis is supported by some dynamical calculations \cite{XLargeSystems, thermalizationJakobIgor1997}, although deviations have also been pointed out \cite{nonEquilibriumBondorf2001} in others.
In spite of their limitations, these approaches have provided important insights into many aspects of the multifragment emission process \cite{AMDReview2004,timeScaleDynStatistical2006,FMD2000,BonaseraBUU,AichelinPhysRep,BurgioSpinodalInstabilities1994,constrainedMolecularDynamics, nonEquilibriumBondorf2001,thermalizationJakobIgor1997,XLargeSystems,ReviewIgor2006,BettyPhysRep2005,GrossPhysRep,Bondorf1995}, although a clear scenario of the phenomenon has yet to emerge, as some of the models draw conflicting pictures of it.
For instance, the results reported in Refs. \cite{AMDReview2004,timeScaleDynStatistical2006} suggest that the deexcitation of the primary fragments takes place concomitantly with their creation, whereas a two-stage scenario, considering a prompt breakup followed by the deexcitation of the primordial hot fragments, has proven to be quite successful in describing many experimental observations \cite{ReviewIgor2006,BettyPhysRep2005,GrossPhysRep,Bondorf1995,smmWeisskopf2018,ISMMlong}.

The situation from the experimental point of view is perhaps even more complicated, as they detect the reaction products after a time span of several orders of magnitude larger than the time scale associated with the primary fragment emission \cite{BondorfEarlyFragRecognition,mechanismsIMF1998,BorderiePhaseTransition2008,CoulombChronometry2015,timeScaleBoulgault1989,
timeScaleLouvel1994,timeScaleExpansionEmission1995,timeScale2000,timeScale2001}.
Owing to this fact, most of the quantities observed experimentally are strongly skewed by the decay of the hot fragments produced in the stages one intends to investigate \cite{ReviewIgor2006,Bondorf1995,smmWeisskopf2018,ISMMlong,IsospinSymmetry}.
Therefore, this fact should be taken into account in comparisons between model calculations and experimental observations.
This adds further uncertainties to the understanding of the problem as it requires additional modelling, although there are some robust observables which are expected to be weakly affected by the primordial fragments' deexcitation \cite{isoscaling3,isoSMMTF}.
Despite such difficulties, experimental studies have provided much important information, which has helped establish more accurate representations for the actual scenario of the multifragmentation process \cite{BorderiePhaseTransition2008,BettyPhysRep2005,reviewSubal2001}.
Furthermore, a few groups have recently developed techniques to reconstruct information corresponding to the freeze-out configuration from the final yields \cite{freezeOut,primFragsIndra2003,primaryFrags2014,primaryFrags2014_2,expRecEex2013,fragReconstructionNPA}.

Studies based on nuclear interferometry techniques have played a very important role in determining some properties of the decaying source \cite{reviewInterferometryKonrad1990,interferometryKoonin,interferometryPratt1991,interferometryBetty1987}.
Indeed, many works \cite{reviewInterferometryKonrad1990,interferometryKoonin,interferometryPratt1991,interferometryBetty1987} demonstrated that such analyses furnish information related to the size of the disassembling system as well as to the time scale associated with particle emission.
However, contributions associated with the fast emission of particles originating in the hot phase of the process and others from the slow component associated with later stages may lead to ambiguities and inaccurate conclusions.
This point has been addressed in recent studies that pointed out means to disentangle information associated with the fast and the slow components of the two-proton correlation functions \cite{interferometryGiuseppe2002,interferometryGiuseppe2003}. 

The present study follows this line and focuses on the contribution to the two-particle correlation functions built from fragments produced in the prompt breakup of an equilibrated nuclear source, as well as in their deexcitation.
We employ a hybrid model recently developed in Ref.\ \cite{smmWeisskopf2018} in which the Improved Statistical Multifragmentation Model (ISMM) \cite{ISMMlong} is used to describe the production of the primary fragments.
On an event by event basis, the latter are allowed to propagate under the influence of their mutual Coulomb repulsion as they continuously deexcite while moving away from each other.
Therefore, this treatment allows one to follow the dynamics and investigate the contributions associated with the different stages.
The model is briefly sketched in Sect.\ \ref{sect:model} and, in Sect.\ \ref{sect:results}, it is applied to the two-particle correlation analysis.
The sensitivity of the results to a few ingredients of the model, such as the fusion cross-section employed in the calculation of the fragment deexcitation, is also examined.
We conclude in Sect.\ \ref{sect:conclusions} with a brief summary of our main findings.

\end{section}
 
\begin{section}{Theoretical framework}
\label{sect:model}
As mentioned above, one of the simplified representations to the complex many-body scenario in which fragments are produced consists in assuming that an equilibrated source of mass and atomic numbers $A_0$ and $Z_0$, respectively, is formed at temperature $T$ and breaks up simultaneously into hot fragments and nucleons.
The former subsequently deexcite emitting lighter fragments and nucleons, besides gamma rays (not included in this treatment), while traveling to the detectors.

In the hybrid model developed in Ref.\ \cite{smmWeisskopf2018}, the prompt breakup stage is described by the canonical version \cite{Bondorf1995,smmIsobaric} of the ISMM \cite{ISMMlong}, in which the primary hot fragments are placed inside a breakup volume $V=(1+\chi)V_0$, where $\chi$ is a model parameter fixed in this work to $\chi=2$ and $V_0$ denotes the source's volume at normal density.
Events are generated as follows:

\begin{description}
\item{i-} A fragmentation mode $f$ is sampled among the possible ones and its statistical weight $W_f$ is calculated from the corresponding Helmholtz free-energy.
\item{ii-} The fragments' momenta $\{\vec{p}_i\}$ are assigned according to the Boltzmann distribution, subject to the constraint $\sum_i \vec{p}_i=0$.
Their positions are also sampled inside the breakup volume $V$, with the condition that $\sum_i A_i\vec{r}_i=0$, where $A_i$ stands for the mass of the $i$-th fragment.
The dynamics starts and the fragments' trajectories are calculated using the Runge-Kutta-Cash-Karp method \cite{RungeKuttaCashKarp}.
\item{iii-} Nuclei of mass number larger than 4, as well as alpha particles, are considered as complex structures, having internal degrees of freedom.
Thus the excitation energy $E^*_i$ of the $i$-th fragment of this kind is sampled with probability:
$$
P(E^*_i)\propto \exp\left(-E^*_i/T\right)\rho_i(E^*_i)\;,
$$
\noindent
where $\rho_i(E^*_i)$ denotes its density of states.
\item{iv-} For a given excitation energy, the total decay width $\Gamma(A_i,Z_i,E^*_i)$ is calculated using the Weisskopf-Ewing approach \cite{Weisskopf,smmWeisskopf2018}. From it, the time $t_i$ at which the fragment decays has probability density:
$$
\tilde P(t_i)=\frac{\Gamma(A_i,Z_i,E^*_i)}{\hbar}\exp\left(-t_i\Gamma(A_i,Z_i,E^*_i)/\hbar\right)\;.
$$
\noindent
One should note that t=0 corresponds to the moment the system disassembles. 
\item{v-} During the dynamics, at time $t_i$, one of the possible decay channels $\lambda$ is chosen with probability:
$$
P_\lambda=\frac{\Gamma_\lambda(A_i,Z_i,E^*_i)}{\Gamma(A_i,Z_i,E^*_i)}\;,
$$
\noindent
where $\Gamma_\lambda(A_i,Z_i,E^*_i)$ symbolizes the partial decay width associated with channel $\lambda$.
\item{vi-} Then, the relative kinetic energy $k$ of the products is sampled according to:
$$
P_k\propto k \sigma_{cC}(k)\rho_C(E^*_i-k)\Theta(K_{\rm max}-k)\;,
$$
\noindent
where $c$ ($C$) denotes the emitted particle (daughter nucleus), $K_{\rm max}$ the largest kinetic energy of the decay, $\sigma_{cC}(k)$ represents the fusion cross-section of the inverse reaction $c+C \rightarrow i$, and $\Theta(x)$ is the standard step function.
\item{vii-} The decay time $t_c$ of $C$ is calculated as in iv above and the new particles, $c$ and $C$, are added to the dynamics, which
continues until the next decay takes place.
When this occurs, we proceed from iv to vii.
\item{viii-} The dynamics is followed until the Coulomb energy of the system is negligible and all the fragments in particle unstable states have decayed.
One then obtains the final yields and the asymptotic fragments' momenta.
\end{description}

\noindent
The emission of fragments with mass number $A_c\le 20$ is allowed in the deexcitation process.

A large number of events (in this work $10^8$) are generated following steps i to viii and the average values of observables are calculated through:

$$
\overline{O}=\frac{\sum_f W_f O_f}{\sum_f W_f}\;.
$$

\noindent
Further details on the model are provided in Ref.\ \cite{smmWeisskopf2018}.

\end{section}

\begin{section}{Results}
\label{sect:results}
We consider the decay of the $A_0=168$ and $Z_0=75$ source, which corresponds to 75\% of the $^{112}$Sn + $^{112}$Sn system.
This choice is made assuming that 25\% of the total system is emitted in the pre-equilibrium stages, leaving an equilibrated source corresponding to 75\% of the colliding nuclei.

\begin{figure}[tb]
\includegraphics[width=9cm,angle=0]{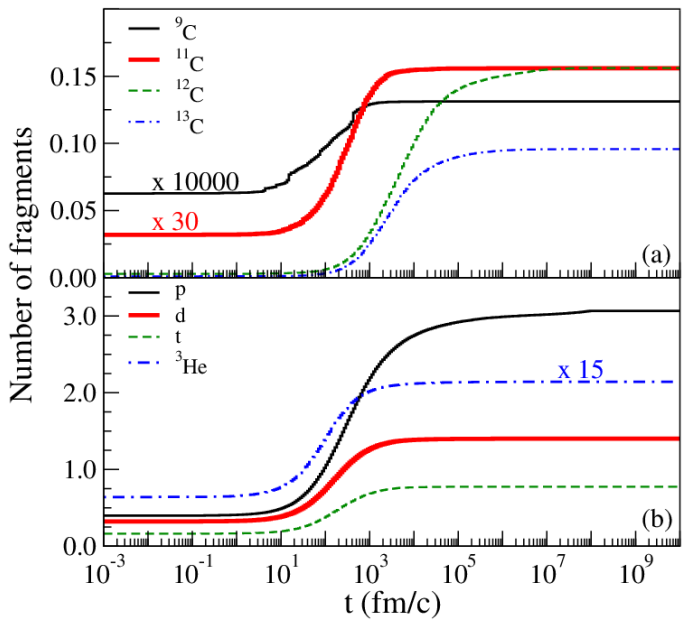}
\caption{\label{fig:popt} (Color online) Number of fragments of different species which are created up to time $t$ and do not undergo subsequent deexcitation, until the asymptotic configuration is reached. For details see the text.}
\end{figure}

We start by examining the number of fragments produced up to time $t$ that survive until the asymptotic configuration is reached.
One notes from Fig.\ \ref{fig:popt} that a small fraction (typically $\approx 10\%$) of the final yields has already been created in the very early stages ($t\lesssim 10$ fm/c) of the process and can be associated with the prompt breakup of the source.
In particular, protons are created during the entire process, beginning with the fragmentation of the source but not exclusive to it.
This suggests that any information they hold concerning the initial fragmenttaion will be contaminated by the post breakup dynamics.
To a lesser extent, this conclusion holds for most species displayed in Fig.\ \ref{fig:popt}.

One should, nevertheless, remark that the production of the neutron deficient carbon isotopes is concentrated in a relatively small time span, starting right after the breakup of the source.
In contrast, the output of the other carbon isotopes becomes important roughly when the population of the neutron deficient ones has reached its asymptotic value.
Although this conclusion might be somewhat sensitive to the treatment employed to describe the fragments' deexcitation, these findings agree qualitatively with the Expanding Evaporating Source (EES) Model which predicts that proton rich fragments are evaporated earlier than proton deficient ones \cite{CoolDyn2006}.
We have also performed calculations with GEMINI++ which agree qualitatively with this conclusion.

\begin{figure}[tb]
\includegraphics[width=9cm,angle=0]{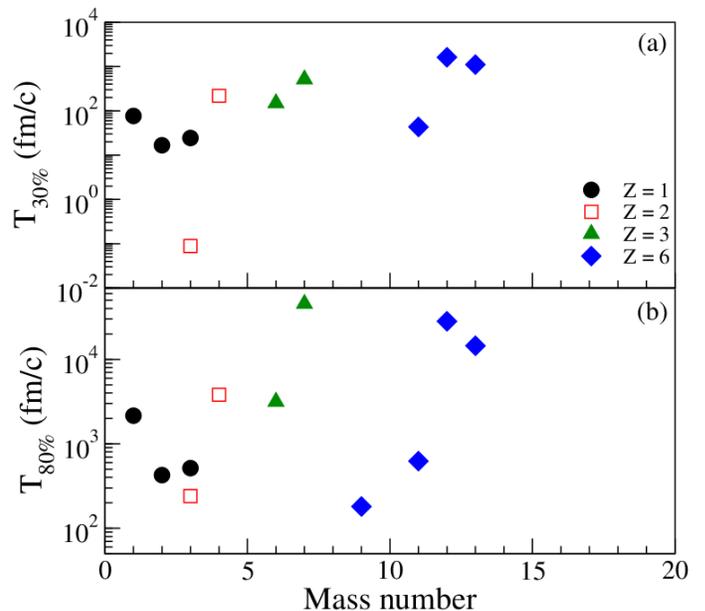}
\caption{\label{fig:dt} (Color online) For different species, time interval for the fragment's multiplicity to attain (a) $30\%$ and (b) 80$\%$ of the asymptotic value. For details see the text.}
\end{figure}

In order to provide more quantitative information on this point, Figs.\ \ref{fig:dt} (a) and (b) respectively exhibit, for some selected species, the time interval necessary for the fragment multiplicity to rise up to 30$\%$ and 80$\%$ of the asymptotic value.
The results reveal that neutron deficient He, Li, and C isotopes are, indeed, associated with shorter time scales than the neutron rich ones.
This is particularly prominent in the case of C isotopes.
One notes that, except for the latter, the final population attains appreciable values only after several hundreds fm/c.
Our results, therefore, suggest that there are isotopes which should be preferentially exploited in nuclear thermometry as they are expected to reflect more trustworthily properties of the system closer to its breakup configuration.
In the same vein, comparison between the results obtained with a transport model and the experimental data reported in Ref.\ \cite{ddCorrelation} suggests that the time span associated with the proton emission is larger than that corresponding to deuterons.
Our model corroborates this conclusion, as may be observed in Figs.\ \ref{fig:popt} and \ref{fig:dt}, which show that protons are continuously produced during the post breakup dynamics.
Their emission is a very important channel over a broad range of excitation energies.

Next, we examine the correlation between two fragments of momentum $\vec{p}_1$ and $\vec{p}_2$, which is defined experimentally as \cite{reviewInterferometryKonrad1990,interferometryGiuseppe2002,interferometryGiuseppe2003}:

\begin{equation}
1+R(\vec{Q},\vec{q})=C\frac{Y_{1,2}(\vec{p}_1,\vec{p}_2)}{Y_1(\vec{p}_1)Y_2(\vec{p}_2)}\;,
\label{ppce}
\end{equation}

\noindent
where $C$ is a normalization constant conveniently chosen so that $<R(\vec{Q},\vec{q})>\rightarrow 0$ at large relative momentum $\vec{q}=\overline{A}(\vec{p}_1/A_1-\vec{p}_2/A_2)$.
In the previous expressions, $\overline{A}=A_1A_2/(A_1+A_2)$, $\vec{Q}=\vec{p}_1+\vec{p}_2$, $Y_i(\vec{p})$ denotes the multiplicity of the species $i$ with momentum $\vec{p}$ and
$Y_{1,2}(\vec{p}_1,\vec{p}_2)$ represents the two particle coincidence yield of the selected species, with momenta $\vec{p}_1$ and $\vec{p}_2$.

\begin{figure}[tb]
\includegraphics[width=9cm,angle=0]{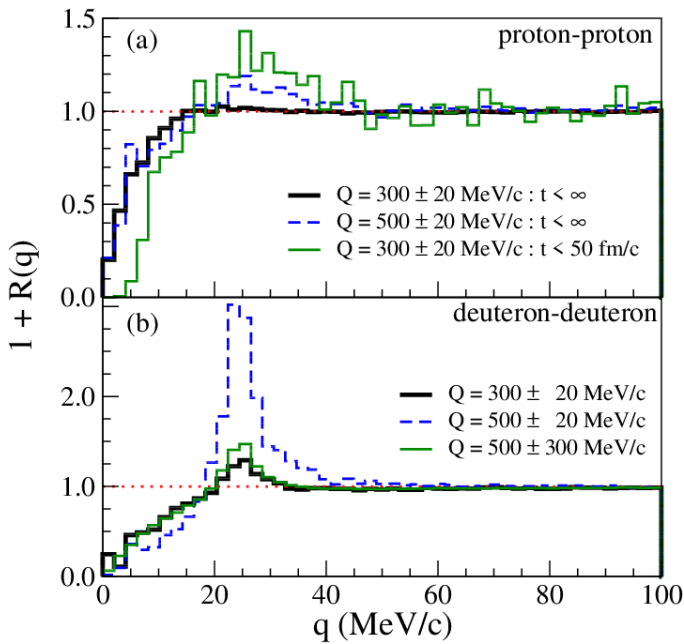}
\caption{\label{fig:ppdd} (Color online) (a) Two-proton and (b) two-deuteron correlation functions calculated using different average $Q$ values.
In frame (a) fragments created within different time spans after the breakup of the source are considered and in frame (b) different $Q$ acceptances are used. For details see the text.}
\end{figure}

\noindent
Theoretically, the correlation function can be calculated using the Pratt-Koonin formula \cite{interferometryKoonin,interferometryPratt1991}

\begin{widetext}
\begin{equation}
1+R(\vec{Q},\vec{q})=\frac{\int dt_1\, d^3\vec{r}_1\, dt_2\, d^3\vec{r}_2\, \Pi_1(\vec{p}_1,t_1,\vec{r}_1)\Pi_2(\vec{p}_2,t_2,\vec{r}_2)
\mid\phi[\vec{q},\vec{r}_1-\vec{r}_2-(t_2-t_1)\vec{Q}/(m_1+m_2)]\mid^2}
{\int dt_1\, d^3\vec{r}_1\, dt_2\, d^3\vec{r}_2\, \Pi_1(\vec{p}_1,t_1,\vec{r}_1)\Pi_2(\vec{p}_2,t_2,\vec{r}_2)}\;,
\label{eq:ppct}
\end{equation}
\end{widetext}

\noindent
where $\Pi_i(\vec{p},t,\vec{r})$ denotes the probability of creating a particle of species $i$ with momentum $\vec{p}$, at time $t$, at position $\vec{r}$, and
$\phi$ symbolizes the scattering wave function of the pair of particles with relative momentum $q$.
As described in Ref.\cite{interferometryKoonin}, the wave function $\phi$ is calculated including contributions of the nuclear potential up to p-waves, using
the Reid soft potential \cite{ReidPotential} in the case of the proton-proton system.
In Ref.\ \cite{reviewInterferometryKonrad1990}, different Woods-Saxon parameterizations are given for proton-deuteron, deuteron-deuteron, deuteron-alpha, triton-triton, and proton-alpha systems.

Two-proton and two-deuteron correlation functions obtained with our hybrid model are displayed in Fig.\ \ref{fig:ppdd} for different values of $Q$.
One notes that, for $Q=300$ MeV/c, $R(q)$ quickly rises from -1 to 0, exhibiting a tiny bump near $q=20$ MeV/c in the two-proton case, in contrast to the two-deuteron correlation function which exhibits a noticeable peak around $q\approx 25$ MeV/c.
In both cases, the height of the peak increases substantially if pairs are selected subject to the constraint $Q=500$ MeV/c and acceptance $\pm 20$ MeV/c.
Figure \ref{fig:ppdd}(a) also reveals that the peak is further enhanced if only fragments produced within 50 fm/c after the breakup of the source are considered.
This sensitivity shows that the modelling of the post breakup dynamics may be improved from such studies.
On the other hand, Fig.\ \ref{fig:ppdd}(b) also shows that the enhancement is almost completely suppressed if large $Q$ acceptances are employed, in order to increase the statistics within each $q$ bin.
One should, therefore, keep in mind that large $Q$ acceptances may obscure important features of this analysis.

\begin{figure}[tb]
\includegraphics[width=9cm,angle=0]{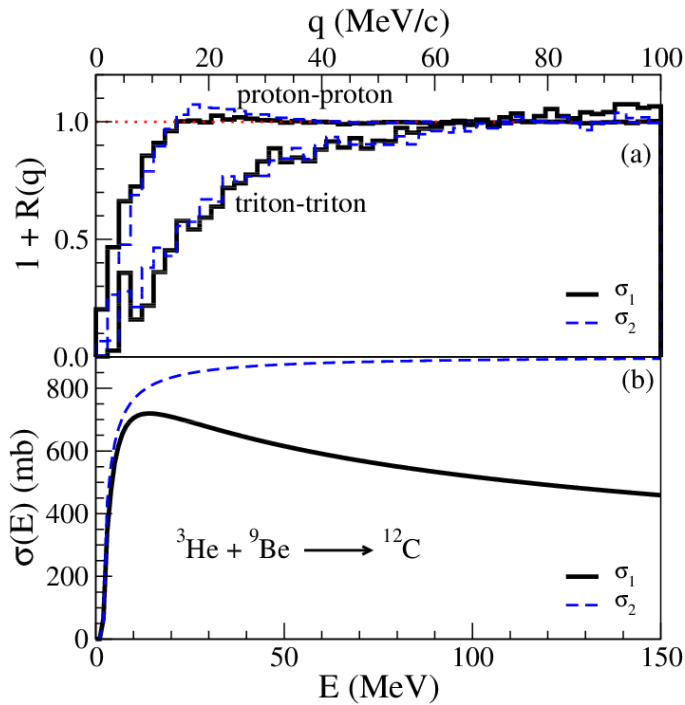}
\caption{\label{fig:ccs} (Color online) (a) Two-proton and two-triton correlation functions obtained with the hybrid-model described in sect.\ \ref{sect:model}, using the fusion cross-sections adopted in Ref.\ \cite{smmWeisskopf2018}, $\sigma_1$, and that employed in Ref.\ \cite{grandCanonicalBotvina1987}, $\sigma_2$.(b) Comparison between these two fusion cross-sections for the reaction $^3{\rm He}\,+\,^9{\rm Be}\rightarrow {^{12}{\rm C}}$. 
For details see the text.}
\end{figure}

Since the fusion cross-section employed in the calculation of the partial decay widths directly affects the emission rates, we investigate the sensitivity of our model calculations to this ingredient.
Figure \ref{fig:ccs}(a) displays the two-proton and the two-triton correlation functions obtained using very different parameterizations for the fusion cross-section.
One of them (denoted here by $\sigma_1$) is the standard one employed in the present hybrid model \cite{smmWeisskopf2018}, whereas the other (labelled $\sigma_2$) is commonly used in SMM calculations \cite{grandCanonicalBotvina1987}.
In order to illustrate the quantitative differences between the two parameterizations, the cross-sections for the $^3{\rm He}\,+\,^9{\rm Be}\rightarrow {^{12}{\rm C}}$ process, given by these prescriptions, are exhibited in Fig.\ \ref{fig:ccs}(b).
One notes that, in spite of the important differences observed in the latter figure, the two-particle correlation functions shown in Fig.\ \ref{fig:ccs}(a) are very similar.
A small peak appears only in the two-proton case.
We have checked that the similarity observed in the two-triton correlation function also occurs in the case  of other light pairs.
Therefore, our results suggest that information on the appropriate parameterization for the inverse reaction cross-section used in such evaporation models is difficult to be obtained from this analysis.

Finally, we  investigate whether the total momentum $\vec{Q}=\vec{p}_1+\vec{p}_2$ of the selected pair may be considered constant during the dynamics.
We therefore evaluate the difference between $\vec{Q}$ at the moment the particles were created $[\vec{Q}(t_0)]$ and  at the asymptotic configuration $[\vec{Q}(t_\infty)]$.
The corresponding distribution, for all deuteron pairs which fulfill the acceptance criterion, $Q=500\pm 20$ MeV/c, is displayed in Fig.\ \ref{fig:dp}(a).
The results show that a very small fraction of the pairs keep their total momentum unchanged during the post breakup dynamics.
Furthermore, the changes are non-negligible when compared to the total momentum.
In order to examine the extent to which $R(q)$ is sensitive to this property of the fragments' dynamics, we plot, in Fig.\ \ref{fig:dp}(b), the two-deuteron correlation function calculated through Eq.\ (\ref{eq:ppct}) using the particles' momenta at $t_0$ (full line) and the asymptotic ones (dashed line).
One sees that the pronounced bump associated with the former case is almost entirely suppressed  in the latter, which exhibits a rather flat behavior. 
We have checked that, in the two-proton case, approximately 20\% of the pairs keep their momenta unchanged during the dynamics and that $R(q)$ is fairly insensitive to it.
It thus suggests that this point should be further investigated in order to determine the extent to which the limitations of this analysis may affect conclusions drawn from different pairs of particles employed in the calculation of $R(q)$.

\begin{figure}[tb]
\begin{tabular}{c}
\includegraphics[width=8cm,angle=0]{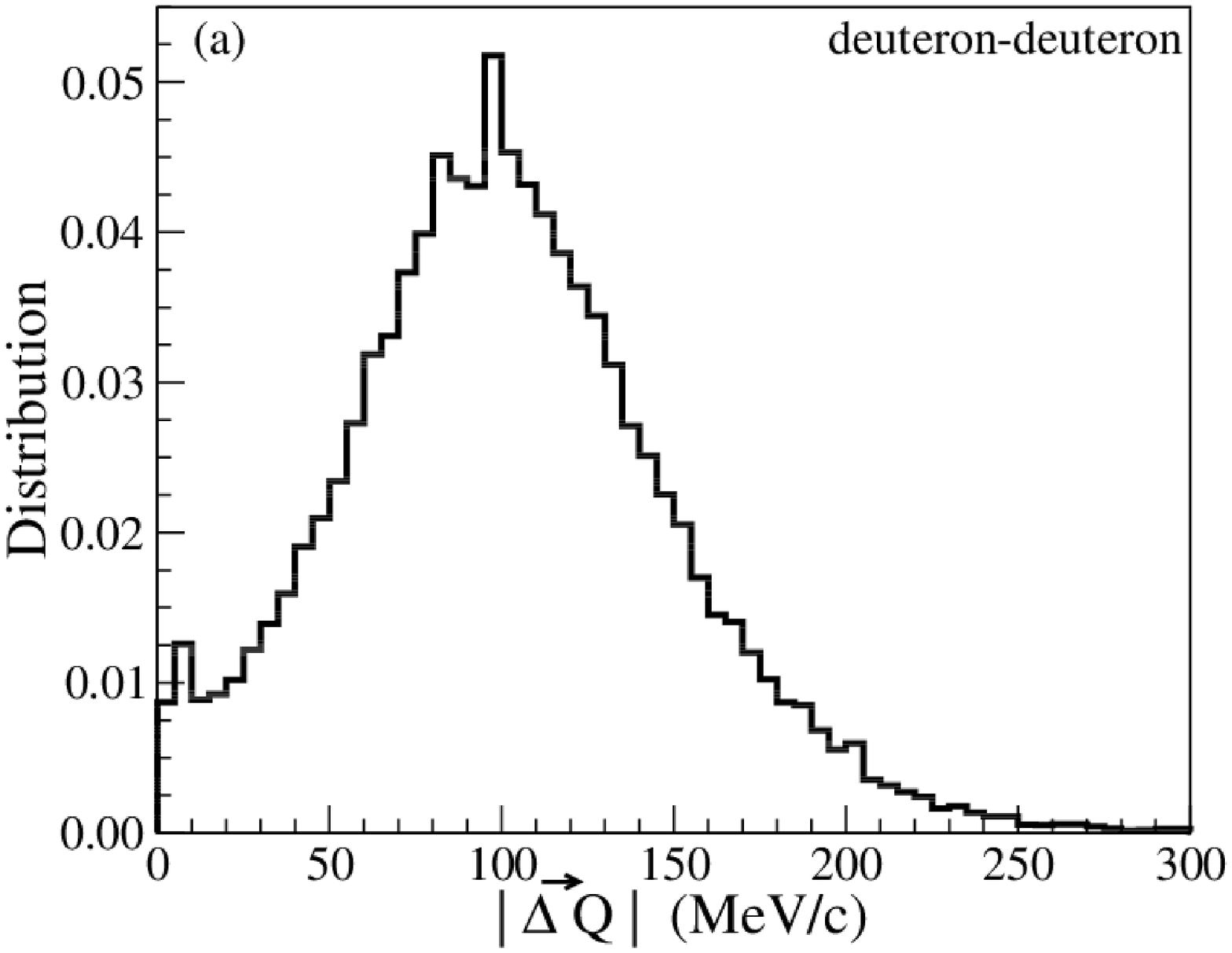}\\
\includegraphics[width=8cm,angle=0]{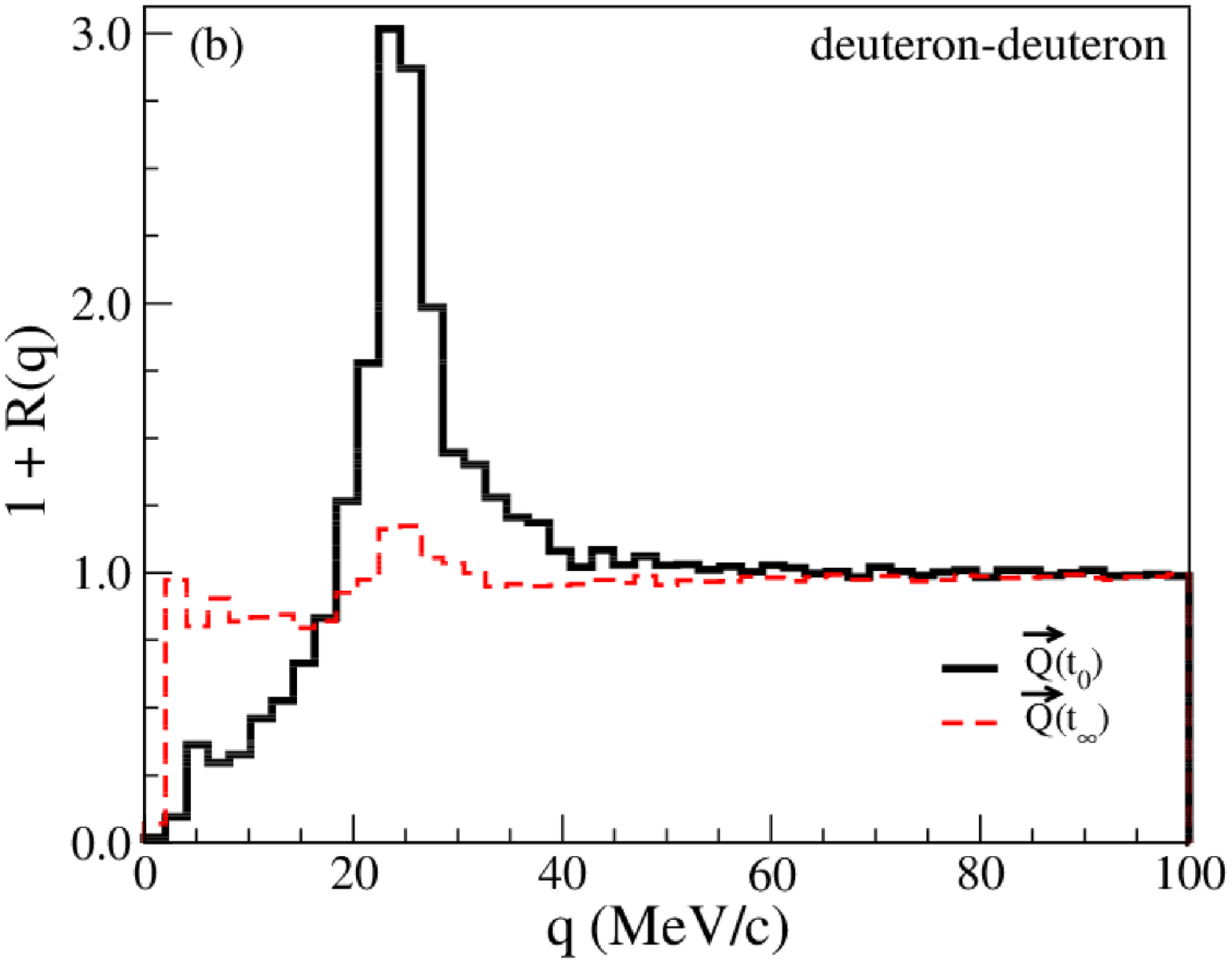}
\end{tabular}
\caption{\label{fig:dp} (Color online) (a) Distribution of $\mid\Delta(\vec{Q})\mid$ for deuteron-deuteron pairs of total momentum $\vec{Q}$, calculated at the moment they were created ($t_0$), which fulfills the requirement $Q=500\pm 20$ MeV/c.
(b) Two-deuteron correlation function calculated using the particles' momenta values at $t_0$ (full line) and the asymptotic ones, $t_\infty$, (dashed line).
For details see the text.}
\end{figure}

\end{section}

\begin{section}{Concluding Remarks}
\label{sect:conclusions}
We have investigated the post breakup dynamics of an excited nuclear source using a hybrid model, recently developed in Ref.\ \cite{smmWeisskopf2018}.
This model allows one to follow the dynamics of the fragments on an event by event basis, as they deexcite while separating under the influence of their mutual Coulomb repulsion.
Therefore it is suitable to study two-particle correlations.
We have investigated the time span associated with the production of different fragments and found that neutron deficient carbon isotopes are produced on much shorter time scales than neutron rich ones.
This makes them particularly suited for the study of the breakup stage, as they hold information more closely related to this phase of the process, in contrast to the others, which are created much later and keep being produced over much longer periods.
Our results suggest that proton rich fragments should share this feature, although this property is more pronounced in the case of C isotopes.
We have also studied the Pratt-Koonin two-particle correlation function for several pairs of particles.
Our results agree with previous calculations that show that this observable is sensitive to the size of the source and its decay rate, although it is difficult to single out, from such analyses, the best parameterization for the fusion cross-section employed in the calculation of the particle decay widths.
We also found that lenient momentum acceptances employed in the selection of pairs of particles may blur important fingerprints that appear when stricter acceptances are imposed.
Another important finding is that the total momentum of the pair of particles $\vec{Q}=\vec{p}_1+\vec{p}_2$ does not remain constant during the post breakup dynamics.
Instead, we found that the difference between the asymptotic momentum and that at $t_0$, when the particles are created,  exhibits a broad bell shaped distribution, whose average value is non-negligible compared to $\mid \vec{p}_1+\vec{p}_2\mid$ at $t_0$.
In the case of the two-deuteron correlation function, we found its properties to be distorted by this fact, whereas it only affects the two-proton correlation function weakly.
We therefore suggest that this aspects need to be further examined.

\end{section}

\begin{acknowledgments}
This work was supported in part by the Brazilian
agencies Conselho Nacional de Desenvolvimento Cient\'\i ­fico
e Tecnol\'ogico (CNPq), by the Funda\c c\~ao de Amparo \`a Pesquisa
do Estado de S\~ao Paulo (FAPESP) and by the Funda\c c\~ao Carlos Chagas Filho de
Amparo \`a  Pesquisa do Estado do Rio de Janeiro (FAPERJ),
a BBP grant from the latter. We also thank the
Programa de Desarrollo de las Ciencias B\'asicas (PEDECIBA)
and the Agencia Nacional de Investigaci\'on e Innovaci\'on
(ANII) for partial financial support.
This work has been done as a part of the project INCT-FNA,
Proc. No.464898/2014-5.
We also thank the N\'ucleo Avan\c cado de Computa\c c\~ao de Alto Desempenho (NACAD), Instituto Alberto Luiz Coimbra de P\'os-Gradua\c c\~ao e Pesquisa em Engenharia (COPPE), Universidade Federal do Rio de Janeiro (UFRJ), for the use of the supercomputer Lobo Carneiro, where the calculations have been carried out.
\end{acknowledgments}

\bibliography{manuscript}
\bibliographystyle{apsrev4-1}

\end{document}